\documentclass{ws-ijmpd}
\usepackage[super,compress]{cite}
\usepackage{graphicx}
\usepackage{amsmath}
\usepackage{amssymb}
\usepackage{amsxtra}
\def\Journal#1#2#3#4{{#1} {\bf #2} (#4) #3 }

\def\PLB{{ Phys. Lett.}  B}

\def\PRD{{ Phys. Rev.} D}

\def\GaC{ Gravitation and Cosmology}

\def\JETPL{ JETP Lett.}

\def\CQG{ Class. Quantum Grav.}

\def\IJMPA{{ Int. J. Mod. Phys.}  A}

\def\BWP{ Bled Workshops in Physics}

\def\GeV{\,{\rm GeV}}
\def\TeV{\,{\rm TeV}}

\def\({\left(}
\def\){\right)}

\def\beq{\begin{equation}}
\def\eeq{\end{equation}}
\begin{document}

\markboth{M.Khlopov}
{Collider Probes for Dark Matter}

%
\catchline{}{}{}{}{}
%

\title{Probes for Dark Matter Physics}

\author{Maxim Yu. Khlopov}

\address{National Research Nuclear University MEPhI (Moscow Engineering Physics Institute), 115409 Moscow, Russia\\APC laboratory 10,
rue Alice Domon et Leonie Duquet 75205 Paris Cedex 13, France\\
khlopov@apc.univ-paris7.fr}

\maketitle

\begin{history}
\received{Day Month Year}
\revised{Day Month Year}
\end{history}

\begin{abstract}
The existence of cosmological dark matter is in the bedrock of the modern cosmology. The dark matter is assumed to be nonbaryonic and to consist of new stable particles. Weakly Interacting Massive Particle (WIMP) miracle appeals to search for neutral stable weakly interacting particles in underground experiments by their nuclear recoil and at colliders by missing energy and momentum, which they carry out.
However the lack of WIMP effects in their direct underground searches and at colliders can appeal to other forms of dark matter candidates. These candidates may be weakly interacting slim particles, superweakly interacting particles, or composite dark matter, in which new particles are bound. Their existence should lead to cosmological effects that can find probes in the astrophysical data. However if composite dark matter contains stable electrically charged leptons and quarks bound by ordinary Coulomb interaction in elusive ”dark atoms”, these charged constituents of dark atoms can be the subject of direct experimental test at the colliders. The models, predicting stable particles with charge -2 without stable particles with charges +1 and -1 can avoid
severe constraints on anomalous isotopes of light elements and provide solution for the puzzles of dark matter searches. In such models the excessive -2 charged particles are bound with primordial helium in O-helium ”atoms”, maintaining specific
nuclear-interacting form of the dark matter. The successful development of composite dark matter
scenarios appeals to experimental search for doubly charged constituents of dark atoms, making experimental search for exotic stable double charged particles \textit{experimentum crucis} for dark atoms of composite dark matter.
\end{abstract}

\section{Introduction}\label{intro}
The data of precision cosmology favor inflationary models with baryosynthesis and dark matter/energy for the structure and evolution of the Universe. These bedrocks of the modern cosmology imply physics beyond the Standard model (BSM) of elementary particles. The processes of inflation and baryosynthesis, as well as the origin of dark matter are widely related to the super high energy range, which is not accessible for direct experimental study at particle accelerators. It generally makes probes for physics of the modern cosmology dominantly indirect.

Indeed, new physical phenomena, corresponding to a high energy scale $\Lambda$ can be directly studied only at energies $E\ge \Lambda$, while at much smaller energies $E\ll \Lambda$ their experimental signatures are related with rare processes suppressed by some power of $E/\Lambda$. The existence of natural conditions for super high energy physics, like Ultra High Energy Cosmic rays or very early Universe extends the field of indirect probes for new physics at the expense of the loss of experimental control over its super high energy sources. 
It makes of special interest particle physics of dark matter that is accessible to collider probes. 

Cosmological arguments specify possible properties of
particle dark matter candidates. Such particles (see e.g. Ref. \citen{book, newBook,DMRev,bertone,LSSFW,Gelmini,Aprile:2009zzd,Feng:2010gw} for review and reference) should be stable, provide the measured dark matter density and be decoupled
from plasma and radiation at least before the beginning of matter dominated stage. The easiest way to
satisfy these conditions is to involve neutral elementary weakly interacting massive particles (WIMPs). Their miraculous feature is that at masses of several hundred GeV such particles can not only explain all the observed dark matter density but also provide collider test for their existence.

Supersymmetric (SUSY) models are of particular interest in this aspect predicting together with possible WIMP candidate (e.g. neutralino) a set of new supersymmetric particles accessible to their experimental search at the LHC. 
However in the lack of positive evidence for SUSY particles at the colliders, as well as in view of controversial results of direct WIMP searches in the underground detectors it may not be the only particle physics solution for the dark matter problem.

Here after brief general review of possible experimental probes for particle dark matter candidates we draw special attention to one of
possible solutions, based on the existence of heavy stable charged particles bound in neutral ”dark atoms”. 
\section{Probes for particle dark matter candidates \label{probes}}
Most of the known particles are unstable. For a particle with the
mass $m$ the particle physics time scale is $t \sim 1/m$
\footnote{Here and further, if it isn't specified otherwise we use the units $\hbar=c=k=1$}, so in
particle world we refer to metastable particles with lifetime $\tau \gg 1/m$ as to sufficiently stable. However to be of cosmological significance metastable
particle should survive in the Big Bang Universe after $t \sim (m_{Pl}/m^2)$, when the temperature of the Universe $T$
fell down below $T \sim m$ and particles go out of thermal equilibrium. It means that the particle lifetime
should exceed $t \sim (m_{Pl}/m) \cdot (1/m)$ and such a long
%
%
lifetime should be explained by the existence of a conservation law, reflecting particle
symmetry. From this viewpoint, physics of dark matter is sensitive to the strict or nearly strict fundamental particle symmetries.

\subsection{WIMP miracle}
 \label{WIMPs}
The simplest form of dark matter candidates is the gas of new
stable neutral massive particles, originated from early Universe. Their stability can be protected by some discrete (as R-parity in supersymmetry) or continuous symmetry.

For
particles with the mass $m$, at high temperature $T>m$ the
equilibrium condition, \beq n \cdot \sigma v \cdot t > 1 \label{eqv} \eeq is valid, if
their annihilation cross section $$\sigma > 1/(m m_{Pl})$$ is
sufficiently large to establish the equilibrium. At $T<m$ such
particles go out of equilibrium and their relative concentration
freezes out. 

If particles have mass in the range of tens-hundreds GeV and annihilation cross section corresponding to weak interaction, the primordial frozen out abundance of such Weakly Interacting Massive Particles (WIMPs) can explain the observed dark matter density. This
is the main idea of the so called \textit{WIMP miracle}
 (see e.g. Refs.
\citen{book,newBook,DMRev,DDMRev} for details). 

It was first found in Ref. \citen{ZKKC} that the process of WIMP annihilation to ordinary particles in the Galaxy should contribute to the cosmic ray and gamma ray fluxes. Though such annihilation involves a sparse fraction of WIMPs it can lead to significant effect even for a subdominant WIMP component in the total dark matter density\cite{DKKM} - supporting its indirect searches in cosmic ray experiments. 

The process of WIMP annihilation to ordinary particles, considered in $t$-channel,
determines their scattering cross section on ordinary particles and thus
relates the primordial abundance of WIMPs to their scattering rate in the
ordinary matter. Forming nonluminous massive halo of our Galaxy, WIMPs can penetrate
the terrestrial matter and scatter on nuclei in underground detectors. The strategy of
direct WIMP searches implies detection of recoil nuclei from this scattering.

The process inverse to annihilation of WIMPs corresponds to their production in collisions
of ordinary particles. It should lead to effects of missing mass and energy-momentum,
being the challenge for experimental search for production of dark matter candidates at accelerators,
e.g. at the LHC.
\subsection{Weakly Interacting Slim Particles}
A wide class of particle models possesses a symmetry breaking
pattern, which can be effectively described by
pseudo-Nambu--Goldstone (PNG) field (see Refs. \citen{book2,dmrps} for review and references). The coherent oscillations of this field represent a specific type
of cold dark matter (CDM). In spite of a very small mass of PNG particles $m_a=\Lambda^2/f$, where $f \gg \Lambda$, these particles are created in Bose-Einstein condensate in the ground state, i.e. they are initially created as nonrelativistic in the very early Universe.
This feature, typical for invisible axion models can be the general feature for all the axion-like PNG particles.

At high temperatures the pattern of successive spontaneous and manifest breaking of global U(1) symmetry implies the
succession of second order phase transitions. In the first
transition at $T \sim f$, continuous degeneracy of vacua leads, at scales
exceeding the correlation length, to the formation of topological
defects in the form of a string network; in the second phase
transition at $T \sim \Lambda \ll f$, continuous transitions in space between degenerated
vacua form surfaces: domain walls surrounded by strings. This last
structure is unstable, but, as was shown in the example of the
invisible axion \cite{Sakharov2,kss,kss2}, it is reflected in the
large scale inhomogeneity of distribution of energy density of
coherent PNG (axion) field oscillations. This energy density is
proportional to the initial value of phase, which acquires dynamical
meaning of amplitude of axion field, when axion mass is switched on
in the result of the second phase transition (see e.g. Ref. \citen{kim} and references therein). The axion mass is given by \beq
m_a=C m_{\pi}f_{\pi}/f,\label{axion}\eeq where $m_{\pi}$ and $f_{\pi}\approx m_{\pi}$ are the pion mass and constant, respectively, the constant $C\sim 1$ depends on the choice of the axion model and $f\gg f_{\pi}$ is the scale of the Peccei-Quinn symmetry breaking) 

Axion couplings to quarks and leptons can lead to rare processes like $K \rightarrow \pi a$ or $\mu \rightarrow e a$. In the gauge model of family symmetry breaking the PNG particle called \textit{archion}  shares properties of axion with the ones of singlet Majoron and familon, being related to the mechanism of neutrino mass generation. In this model together with archion decays of quarks and charged leptons archion decays $\nu_H \rightarrow \nu_L a$ of heavier neutrino $\nu_H$ to lighter neutrino $\nu_L$ are also predicted\cite{berezhiani5}. 

The relationship (\ref{axion}) of axion to neutral pion makes possible to estimate the cross section of axion interactions from the corresponding cross section of pion processes multiplied by the factor $(f_{\pi}/f_a)^2$. In particular, in the analogy with $\pi^0$ decay axion decay to two photons is possible. Such axion-photon coupling leads to effect of "light shining through the wall" due to axion-photon conversion in the radio frequency cavity\cite{sikivie} or in strong magnetic fields of astrophysical objects\cite{raffelt}. Axion emission by stars should influence the rate of stellar evolution, putting observational constraint on the axion parameters.

\subsection{Super-Weakly Interacting Massive Particles}
The maximal
temperature, which is reached in inflationary Universe, is the
reheating temperature, $T_{r}$, after inflation. So, the very
weakly interacting particles with the annihilation cross section
$\sigma < 1/(T_{r} m_{Pl}),$ as well as very heavy particles with
the mass $m \gg T_{r}$ can not be in thermal equilibrium, and the
detailed mechanism of their production should be considered to
calculate their primordial abundance.

In particular, thermal production of gravitino in very early Universe is proportional to the reheating temperature $T_{r}$, what puts upper limit on this temperature from constraints on primordial gravitino abundance\cite{khlopovlinde,khlopovlinde2,khlopovlinde3,khlopov3,khlopov31,Karsten,Kawasaki}.

In the minimal Starobinsky-Polonyi N=1 Supergravity superheavy gravitino are produced in decays of inflaton and Polonyi fields. It links physics of inflation to physics of superheavy gravitino dark matter in these models (see Ref. \citen{AKK} in this issue for review and references).
\subsection{Composite Dark Matter}
Extensive hidden sector of particle theory can provide the existence of new interactions, which only new particles possess. It gives rise to Composite Dark Matter, in which dark matter species are composed of constituents, bound by various forces. 

Historically one of the first examples of such composite dark matter was presented by the model of mirror matter. Mirror particles, first proposed by T. D. Lee and C. N. Yang in Ref. \citen{LeeYang} to restore equivalence of left- and right-handed co-ordinate systems in the presence of P- and C- violation in weak interactions, should be strictly symmetric by their properties to their ordinary twins. After discovery of CP-violation it was shown by I. Yu. Kobzarev, L. B. Okun and I. Ya. Pomeranchuk in Ref. \citen{KOP} that mirror partners cannot be associated with antiparticles and should represent a new set of symmetric partners for ordinary quarks and leptons with their own strong, electromagnetic and weak mirror interactions. It means that there should exist mirror quarks, bound in mirror nucleons by mirror QCD forces and mirror atoms, in which mirror nuclei are bound with mirror electrons by mirror electromagnetic interaction \cite{ZKrev,FootVolkas}. If gravity is the only common interaction for ordinary and mirror particles, mirror matter can be present in the Universe in the form of elusive mirror objects, having symmetric properties with ordinary astronomical objects (gas, plasma, stars, planets...), but causing only gravitational effects on the ordinary matter\cite{Blin1,Blin2}.

Even in the absence of any other common interaction except for gravity, the observational data on primordial helium abundance and upper limits on the local dark matter seem to exclude mirror matter, evolving in the Universe in a fully symmetric way in parallel with the ordinary baryonic matter\cite{Carlson,FootVolkasBBN}. The symmetry in cosmological evolution of mirror matter can be broken either by initial conditions\cite{zurabCV,zurab}, or by breaking mirror symmetry in the sets of particles and their interactions as it takes place in the shadow world\cite{shadow,shadow2}, arising in the heterotic string model. We refer to Refs.
\cite{newBook,OkunRev,Paolo} for review of mirror matter and its cosmological probes.

For the general case of the shadow world the detailed analysis becomes rather complicated and more simple examples of dark atoms bound by various forces are possible (see e.g. Ref. \citen{adm} for review and references). 
\section{Dark atoms}
In the extensive list of possible dark matter candidates, going far beyond the examples discussed above dark atoms that consist of stable electrically charged particles are of special interest in view of a richness of their experimental and observational probes. In such atoms  new stable electrically charged particles are bound by ordinary Coulomb interactions (see Refs. \citen{DMRev,mpla,DDMRev} and references therein). De-excitation of such atoms should result in electromagnetic radiation or electron-positron pair production, while search for free constituents can be challenging for experimental probes by charged particle detectors\cite{2cBled}.

Stable particles with charge -1 are excluded due to overproduction of anomalous isotopes.  However, there doesn't appear such an evident contradiction for negatively doubly charged particles. If metastable, in a~tree approximation, such particles cannot decay to pair of quarks due to electric charge conservation and only decays to the~same sign leptons are possible. The~latter implies lepton number nonconservation, being a~profound signature of new physics. In general, it makes such states sufficiently long-living in a~cosmological scale.  
\subsection{Charged constituents of dark atoms}
There
exist several types of particle models where heavy
stable -2  charged species, $O^{--}$, are predicted:
\begin{itemize}
\item[(a)] AC-leptons, predicted
as an extension of the Standard Model, based on the approach
of almost-commutative geometry \cite{Khlopov:2006dk,5,FKS,bookAC}.
\item[(b)] Technileptons and
anti-technibaryons in the framework of Walking Technicolor
(WTC) \cite{KK,Sannino:2004qp,Hong:2004td,Dietrich:2005jn,Dietrich:2005wk,Gudnason:2006ug,Gudnason:2006yj}.
\item[(c)] stable "heavy quark clusters" $\bar U \bar U \bar U$ formed by anti-$U$ quark of 4th generation
\cite{Khlopov:2006dk,Q,I,lom,KPS06,Belotsky:2008se} \item[(d)] and, finally, stable charged
clusters $\bar u_5 \bar u_5 \bar u_5$ of (anti)quarks $\bar u_5$ of
5th family can follow from the approach, unifying spins and charges\cite{Norma}.
\end{itemize}
All these models also
predict corresponding +2 charge particles. If these positively charged particles remain free in the early Universe,
they can recombine with ordinary electrons in anomalous helium, which is strongly constrained in the
terrestrial matter. Therefore a cosmological scenario should provide a  mechanism which suppresses anomalous helium.
There are  two possible mechanisms than can provide a suppression:
\begin{itemize}
\item[(i)] The abundance of anomalous helium in the Galaxy may be significant, but in terrestrial matter
 a recombination mechanism could suppress this abundance below experimental upper limits \cite{Khlopov:2006dk,FKS}.
The existence of a new U(1) gauge symmetry, causing new Coulomb-like long range interactions between charged dark matter particles, is crucial for this mechanism. This leads inevitably to the existence of dark radiation in the form of hidden photons.
\item[(ii)] Free positively charged particles are already suppressed in the early Universe and the abundance
of anomalous helium in the Galaxy is negligible \cite{mpla,I}.
\end{itemize}
These two possibilities correspond to two different cosmological scenarios of dark atoms. The first one is
realized in the scenario with AC leptons, forming neutral AC atoms \cite{FKS}.
The second assumes a charge asymmetry  of the $O^{--}$ which forms the atom-like states with
primordial helium \cite{mpla,I}.

If new stable species belong to non-trivial representations of
the SU(2) electroweak group, sphaleron transitions at high temperatures
can provide the relation between baryon asymmetry and excess of
-2 charge stable species, as it was demonstrated in the case of WTC
\cite{KK,KK2,unesco,iwara}.
\subsection{Dark atom cosmology}
 After it is formed
in the Standard Big Bang Nucleosynthesis (BBN), $^4He$ screens the
$O^{--}$ charged particles in composite $(^4He^{++}O^{--})$ {\it
$OHe$} ``atoms'' \cite{I}.
In all the models of $OHe$, $O^{--}$ behaves either as a lepton or
as a specific ``heavy quark cluster" with strongly suppressed hadronic
interactions. 

If $O^{--}$ excess can explain the observed dark matter density, 
the cosmological scenario of the $OHe$ Universe is possible. It involves only one parameter
of new physics $-$ the mass of O$^{--}$. Such a scenario is insensitive to the properties of $O^{--}$ (except for its mass), since the main features of the $OHe$ dark atoms are determined by their nuclear interacting helium shell. 

$OHe$ gas has nuclear interaction with baryons, which are dominantly protons after Big Bang Nucleosynthesis. Before the end of radiation domination stage the rate of expansion exceeds the rate of energy and momentum transfer from plasma to $OHe$ gas and the latter decouples from plasma and radiation. Then $OHe$ starts to dominate at the Matter dominated stage, playing the role of Warmer than Cold Dark Matter in the process of Large Scale Structure formation\cite{mpla,I}. This feature is due to suppression of small scale fluctuations that convert in acoustic waves before $OHe$ decoupling. However, this suppression is not as strong as the suppression of such fluctuations in Warm Dark matter models due to free streaming of few keV dark matter particles.

In terrestrial matter such dark matter species are slowed down and cannot cause significant nuclear recoil in the underground detectors, making them elusive in direct WIMP search experiments (where detection is based on nuclear recoil) such as CDMS, XENON100 and LUX. The positive results of DAMA experiments (see \cite{DAMAtalk} for review and references) can find in this scenario a nontrivial explanation due to a low energy radiative capture of $OHe$ by intermediate mass nuclei~\cite{mpla,DMRev,DDMRev}. This explains the negative results of the XENON100 and LUX experiments. The rate of this capture is
proportional to the temperature: this leads to a suppression of this effect in cryogenic
detectors, such as CDMS. 

OHe collisions in the central part of the Galaxy lead to OHe
excitations, and de-excitations with pair production in E0 transitions can explain the
excess of the positron-annihilation line, observed by INTEGRAL in the galactic bulge \cite{DMRev,DDMRev,KK2,CKWahe}.

One should note that the nuclear physics of OHe is in the course of development and its basic element for a successful and self-consistent OHe dark matter scenario is related to the existence of a dipole Coulomb barrier, arising in the process of OHe-nucleus interaction and providing the dominance of elastic collisions of OHe with nuclei. This problem is the main open question of composite dark matter, which implies correct quantum mechanical solution \cite{CKW}. The lack of such a barrier and essential contribution of inelastic OHe-nucleus processes seem to lead to inevitable overproduction of anomalous isotopes \cite{CKW2}.

\section{Indirect effects of composite dark matter}
\label{astro}
The existence of O-helium should lead to a number of astrophysical signatures,
which can constrain or prove this hypothesis. One of the signatures of O-helium can be a presence
of an anomalous low Z/A component in the cosmic ray flux. O-helium atoms that are present in the
Galaxy in the form of the dark matter can be destroyed in astrophysical processes and free $X^{−−}$ can be
accelerated as ordinary charged particles. O-helium atoms can be ionized due to nuclear interaction with
cosmic rays or in the front of a shock wave in the Supernova explosions, where they were effectively
accumulated during star evolution \cite{I}. If the mechanisms of $X^{−−}$ acceleration are effective, the
low Z/A component with charge −2 should be present in cosmic rays at the level of $F_X/F_p \sim 10^{-9}m^{-1}_o $ \cite{KK2},
and might be observed by PAMELA and AMS02 cosmic ray experiments. Here $m_o$ is the mass
of O-helium in TeV, $F_X$ and $F_p$ are the fluxes of $X^{−−}$ and protons, respectively.
\subsection{Excess of positron annihilation line in the galactic bulge}
\label{bulge}
\begin{figure}[htbp]
    \begin{center}
        \includegraphics[scale=1]{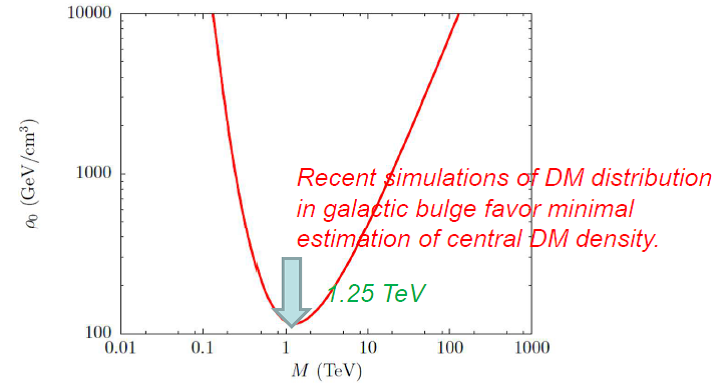}
        \caption{Dark matter is subdominant in the central part of Galaxy. It strongly suppresses it dynamical effect and causes large uncertainty in dark matter density and velocity distribution. At this uncertainty one can explain the positron line excess, observed by INTERGRAL, for a wide range of $m_o$ given by the curve with minimum at $m_o = 1.25 \TeV$. However, recent analysis of possible dark matter distribution in the galactic bulge favor minimal value of its central density.}
        \label{integral}
    \end{center}
\end{figure}

Another signature of O-helium in the Galaxy is the excess of the positron annihilation line in cosmic
gamma radiation due to de-excitation of the O-helium after its interaction in the interstellar space. If
2S level of O-helium is excited, its direct one-photon transition to the 1S ground state is forbidden and
the de-excitation mainly goes through direct pair production. In principle this mechanism of positron
production can explain the excess in positron annihilation line from the galactic bulge, measured by the
INTEGRAL experiment. Due to the large uncertainty of DM distribution in the galactic bulge this interpretation of the INTEGRAL data is possible in a wide range of masses of
O-helium with the minimal required central density of O-helium dark matter at $m_o = 1.25 \TeV$ \cite{integral1,integral2}
For smaller or larger values of $m_o$ one needs larger central density to provide effective excitation of O-helium in collisions.
Current analysis favors lowest values of central dark matter density, making possible O-helium explanation for this excess only for a narrow window around this minimal value (see Fig. \ref{integral} taken from Ref. \citen{2cBled})

\subsection{Composite dark matter solution for high energy positron excess}
\label{HEpositrons}
\begin{figure}[htbp]
    \begin{center}
        \includegraphics[scale=1]{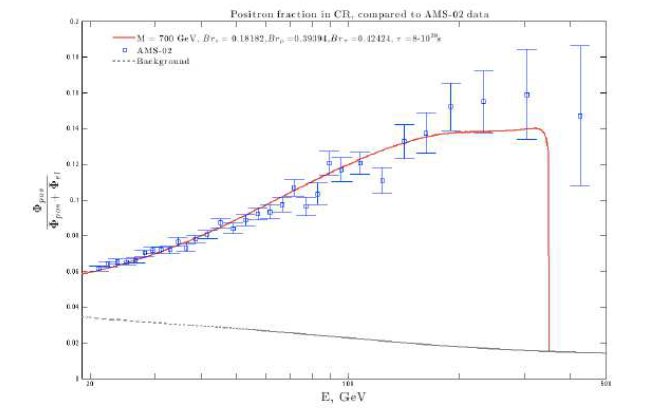}
        \caption{Best fit high energy positron fluxes from decaying composite dark matter in confrontation with the results of AMS02 experiment.}
        \label{ams}
    \end{center}
\end{figure}

In a two-component dark atom model, based on Walking Technicolor, a
sparse WIMP-like component of atom-like state, made of positive and negative doubly charged techniparticles, is present together with the dominant OHe dark atom and the decays of doubly positive charged techniparticles to pairs of same-sign leptons can explain the excess of high-energy cosmic-ray
positrons, found in PAMELA and AMS02 experiments\cite{PAMELA,AMS-2old,AMS2,AMS1}. This explanation is possible for the mass of decaying +2 charged particle below 1 TeV
and depends on the branching ratios of leptonic channels (See Fig. \ref{ams} taken from Ref. \citen{2cBled}).

\begin{figure}[htbp]
    \begin{center}
        \includegraphics[scale=1.2]{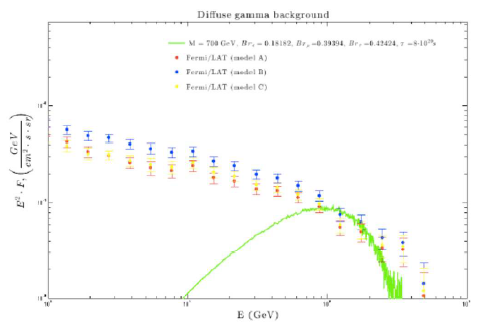}
        \caption{Gamma ray flux accompanying the best fit high energy positron fluxes from decaying composite dark matter reproducing the results of AMS02 experiment, in confrontation with FERMI/LAT measurement of gamma ray background.}
        \label{fermi}
    \end{center}
\end{figure}

Since even pure
lepton decay channels are inevitably accompanied by gamma radiation the
important constraint on this model follows from the measurement of cosmic
gamma ray background in FERMI/LAT experiment\cite{FERMI}. The corresponding effect is shown on Fig. \ref{fermi}, taken from Ref. \citen{2cBled}.
The multi-parameter
analysis of decaying dark atom constituent model determines the maximal model independent value of the mass of decaying
+2 charge particle, at which this explanation is possible $$m_o<1 TeV.$$

One should take into account that even in this range hypothesis on decaying composite dark matter, distributed in the galactic halo, can lead according to Ref. \citen{kb} to gamma ray flux exceeding the measured by FERMI/LAT. It can make more attractive interpretation of these data by an astrophysical pulsar local source\cite{semikoz} or by some local source of dark matter annihilation or decay. With this precaution we turn further following Ref. \citen{2cBled} to the collider probes for composite dark matter constituents.

\subsection{Sensitivity of indirect effect of composite dark matter to the mass of their double charged constituents}
\label{mass}
We see that indirect effects of composite dark matter strongly depend on the mass of its double charged constituents. 

To explain the excess of positron annihilation line in the galactic bulge mass of double charged constituent of O-helium should be in a narrow window around
$$m_o = 1.25 \TeV.$$

To explain the excess of high energy cosmic ray positrons by decays of constituents of composite dark matter with charge +2 and to avoid overproduction of gamma background, accompanying such decays, the mass of such constituent should be in the range
$$m_o < 1 \TeV.$$

These predictions should be confronted with the experimental data on the accelerator search for stable double charged particles.
\section{Collider probes for dark atom constituents}
\label{experiment}
A~new charged massive particle with electric charge $\neq 1e$ would represent a~dramatic deviation from the~predictions of the~Standard Model, and such a~spectacular discovery would lead to fundamental insights and critical theoretical developments. Searches for such kind of particles were carried out in many cosmic ray and collider experiments (see for instance review in~\cite{fair}). 
Experimental search for double charged particles is of a~special interest because of its important probe for dark atom cosmology\cite{2cBled}.

Here we present the~results and prospects for searches for the~multi-charged particles in the~ATLAS and the~CMS collaborations at the LHC, following the review Ref. \citen{2cBled}. 

\subsection{ATLAS experiment at LHC}
\label{ATLAS_search}
In Run~1 (2010--2012), the~ATLAS~\cite{Aad:2008zzm} collaboration at LHC performed two searches for long-lived multi-charged particles, including the~double charged particles: 
one search with $4.4$~fb$^{-1}$ of data collected in $pp$ collisions at $\sqrt{s}=7$~\TeV{}~\cite{Aad:2013pqd}, 
and another one with $20.3$~fb$^{-1}$ collected at $\sqrt{s}=8$~\TeV{}~\cite{Aad:2015oga}. 

Both these searches feature particles with large transverse momentum values, traversing the~entire ATLAS detector. An~energy loss of a~double charged particle is by a~factor of $q^2=4$ higher than that of single charged particle. Such particles will leave a~very characteristic signature of high ionization in the~detector. 
More specifically, the~searches look for particles with anomalously high ionization on their tracks in three independent detector subsystems: silicon pixel detector (Pixel) and transition radiation tracker (TRT) in the~ATLAS inner detector, and monitoring drift tubes (MDT) in the~muon system.

\begin{figure}[htbp]
    \begin{center}
        \includegraphics[scale=0.45]{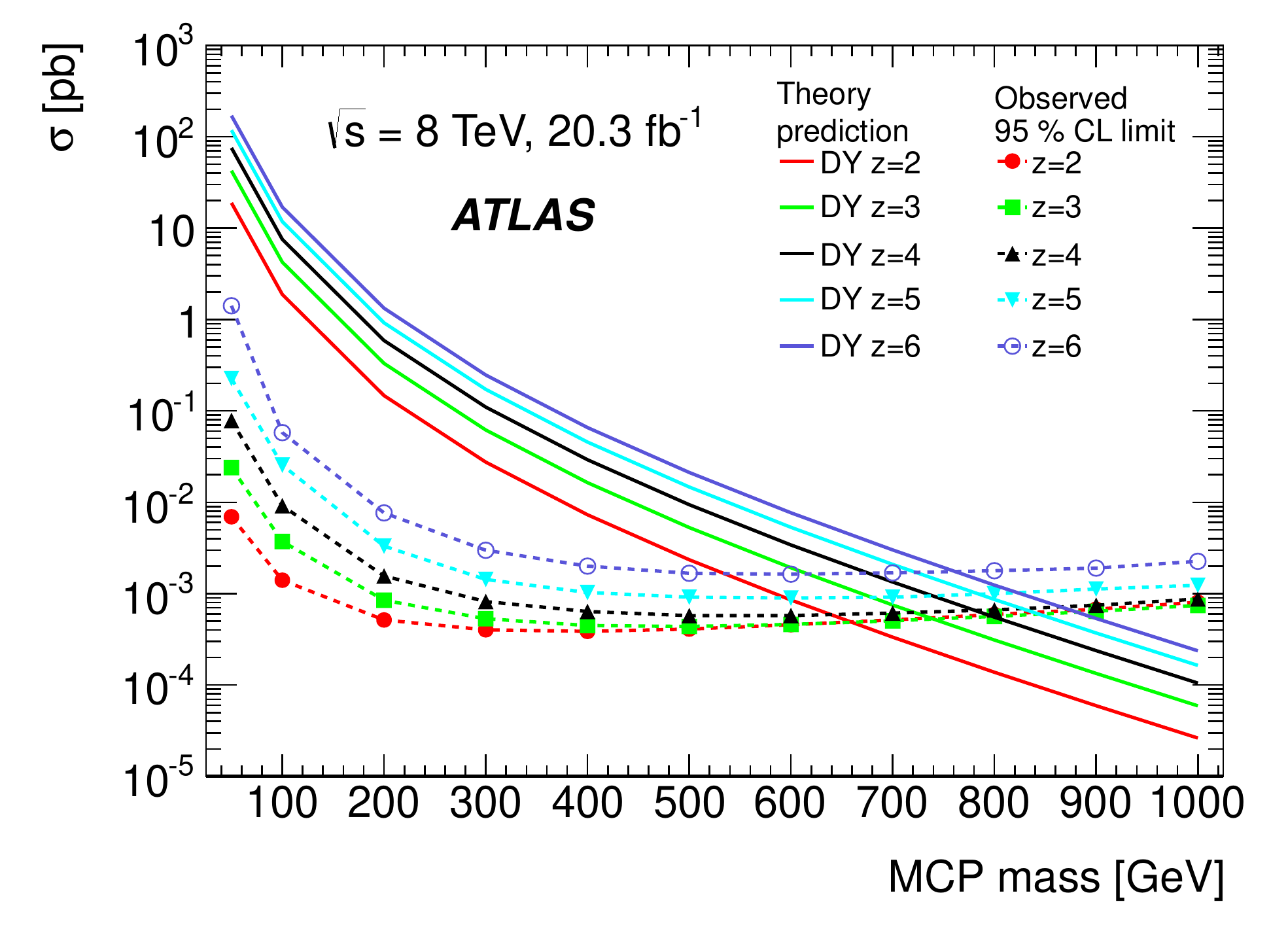}
        \caption{Observed $95\%$ CL cross-section upper limits and theoretical cross-sections as functions of the multi-charged particles mass. Again, the~double charged particles are denoted as ``$z=2$'' (red points and lines). The~picture is taken following Ref.\cite{2cBled} from~Ref.\cite{Aad:2015oga}.}
        \label{ATLAS_Limits}
    \end{center}
\end{figure}
No events with double charged particles were found in neither 2011 or 2012 experimental data sets, setting the~lower mass limits to $430$ and $660$~\GeV{}, respectively, at $95\%$ CL. The~comparison of observed cross-section upper limits and theoretically predicted cross-sections is shown in Fig.~\ref{ATLAS_Limits}.
\subsection {CMS experiment at LHC}
\label{CMS_search}
In parallel to the~ATLAS experiment, the~CMS~\cite{Chatrchyan:2008aa} collaboration at LHC performed a~search for double charged particles, using  $5.0$~fb$^{-1}$ of data collected in $pp$ collisions at $\sqrt{s}=7$~\TeV{} and $18.8$~fb$^{-1}$ collected at $\sqrt{s}=8$~\TeV{}~\cite{Chatrchyan:2013oca}.

For the~$8$~\TeV{} search, the~mass limit of $665$~\GeV{} was obtained. This result (within uncertainties) is very close to the~ATLAS limit of $660$~\GeV{} for the~$8$~\TeV{}  data set. Also, CMS treated the~results obtained at $7$ and $8$~\TeV{} as combined. This allowed to push the~lower mass limit to $685$~\GeV{} at $95\%$ CL. A~combination of the~results of two experiments for $8$~\TeV{} would  mean an~increase of statistics by a~factor of $2$. Having said that, one can conclude that the~mass limit based on the~results of both experiment for double charged particles can be set at the~level of about $730$~\GeV{}.

\subsection {What one expects from LHC Run~2?}
\label{LHC Run 2}

Considering a~recent CMS Physics Analysis Summary~\cite{CMS:2016ybj} and an~ATLAS paper in preparation, both on the~searches for double charged particles in data delivered by LHC to these experiments in 2015--2016, it was concluded in Ref.\citen{2cBled} that each of these two experiments will be able to set a~lower mass limit on the~double charged particles at $m=1000\pm50$~\GeV{}. 
Going further and considering the data taking periods of ATLAS and CMS until the end of Run 2 (end of 2018), a~low limit on the~double charged particles mass corresponding to the~Run 2 data set was estimated in Ref.\citen{2cBled}. 

These estimations showed that the~ATLAS and CMS collaborations may each be expected to set the~lower mass limits at the~level of $1.2$~\TeV{} based on their analyses of the~entire $13$~\TeV{} data set. If these two experiments combined their independently gathered statistics together for this kind of search, the~limits would go as high as up to $1.3$~\TeV{}. It may cover all the range of masses for stable double charged particles, at which the observed excess of low and high energy positrons can be explained as a possible indirect effect of the composite dark matter.
\section{Conclusions}
The existence of heavy stable neutral particles is one of the popular solutions for the dark matter problem. The wide class of such candidates is predicted by particle theory at the super high energy range, making such particles elusive from direct experimental probes.  Therefore the feasibility of dark matter candidates for direct experimental searches is challenging. In the list of such attractive candidates charged constituents of composite are of special interest.

Indeed, DM considered to be electrically neutral, can be formed similar to the ordinary atomic matter by stable heavy charged particles bound in neutral atom-like states by Coulomb attraction. Analysis of the cosmological data and atomic composition of the Universe gives the constrains on the particle charge
showing that only -2 charged constituents, being trapped by primordial helium in neutral O-helium states, can avoid the problem of overproduction of the anomalous isotopes of chemical elements, which
are severely constrained by observations. Cosmological model of O-helium dark matter can even explain
puzzles of direct dark matter searches. The attractive feature of such scenario is the minimal number of involved parameters of new physics. In the wide class of OHe scenario it is the mass of a new stable double charged particle only. One should note, however, that the nontrivial nuclear physics of OHe interaction with matter is still not fully explored in a proper quantum mechanical analysis.

Stable  charge -2 states ($X^{--}$) can be elementary like AC-leptons or technileptons, or look like
technibaryons. The latter, composed of techniquarks, reveal their structure at much higher energy scale
and should be produced at colliders and accelerators as elementary species. They can also be composite like ``heavy quark clusters'' $\bar U \bar U \bar U$ formed by anti-U quark in one of the models of fourth generation \cite{I} or $\bar u_5 \bar u_5 \bar u_5$ of
(anti)quarks $\bar u_5$ of stable 5th family in the approach \cite{Norma}.

In the context of composite dark matter scenario accelerator search for stable doubly charged leptons
acquires the meaning of direct critical test for existence of charged constituents of cosmological dark matter.

The signature for AC leptons and techniparticles is unique and distinctive what allows to separate them from other hypothetical exotic particles. Composite dark matter models can explain observed excess of high energy positrons and gamma radiation in positron annihilation line at the masses of $X^{--}$ in the range of $1$~\TeV{}, it makes search for double charged particles in this range direct experimental test for these predictions of composite dark matter models.

Test for composite $X^{--}$ can be only indirect: through the search for heavy hadrons, composed of single $U$ or $\bar U$ and light quarks (similar to R-hadrons).
 
The~ATLAS and CMS collaborations at the~Large Hadron Collider are searching for the~double charged particles since 2011. The~most stringent results achieved so far exclude the~existence of such particles up to their mass of $680$~\GeV{}. This value was obtained by both ATLAS and CMS collaborations independently. It is expected that if these two collaborations combine their independently gathered statistics of LHC Run 2 (2015--2018), the~lower mass limit of double charged particles could reach the~level of about $1.3$~\TeV{}. It will make search for exotic long-living double charged particles an \textit{experimentum crucis} for interpretation of low and high energy positron anomalies by composite dark matter.
\section*{Acknowledgements}
I express my gratitude to K.Belotsky, O.Bulekov, J.-R. Cudell, C. Kouvaris, M.Laletin, A.Romaniouk and Yu.Smirnov for collaboration in development of the presented approach. The work was supported by Russian Science Foundation and fulfilled in the framework of MEPhI
Academic Excellence Project (contract 02.a03.21.0005, 27.08.2013).


\end{document}